\documentstyle[psfig]{mn}

\newcommand{\apj}{ApJ}
\newcommand{\apjl}{ApJLett}
\newcommand{\araa}{ARA\&A}
\newcommand{\pasp}{PASP}
\newcommand{\aj}{AJ}
\newcommand{\mnras}{MNRAS}
\begin{document}
\title[Gravitational lensing in binary stars]{
Gravitational lensing in eclipsing binary stars}

\author[T.R. Marsh]
{T.R. Marsh\\
Department of Physics and Astronomy, University of Southampton, 
Highfield, Southampton SO17 1BJ}
\date{Accepted ??
      Received ??
      in original form ??}
\pubyear{1999}

\maketitle

\begin{abstract}
I consider the effect of the gravitational deflection of light upon
the light curves of eclipsing binary stars, focussing mainly upon
systems containing at least one white dwarf component. In absolute
terms the effects are small, however they are strongest at the time of
secondary eclipse when the white dwarf transits its companion, and act
to reduce the depth of this feature.  If not accounted for, this may
lead to under-estimation of the radius of the white dwarf compared to
that of its companion. I show that the effect is significant for
plausible binary parameters, and that it leads to $\sim 25$\%
reduction in the transit depth in the system KPD~1930+2752.  The reduction
of eclipse depth is degenerate with the stellar radius ratio, and
therefore cannot be used to establish the existence of lensing. A
second order effect of the light bending is to steepen the ingress and
egress features of the secondary eclipse relative to the primary
eclipse, although it will be difficult to see this in practice. I
consider also binaries containing neutron stars and black-holes. I
conclude that, although relatively large effects are possible in such
systems, a combination of rarity, faintness and intrinsic variability
make it unlikely that lensing will be detectable in them.
\end{abstract}


\section{Introduction}
Gould (1995) considered whether lensing by one star of its companion
within a binary could be a significant source of microlensing events.
He showed that the effect was insignificant except for pairs of
pulsars, although the rarity of these systems probably eliminates them
as well. For white dwarf binaries, Gould concludes that one would need
$10^5$ observations of each of $10^4$ such systems to find a single
example of lensing, a seemingly hopeless task. However, Gould's
discussion was couched in the standard framework of microlensing in
which ``lensing'' is only said to occur if the source lies within the
Einstein radius of the lens, giving at least $0.3$ magnitudes of
amplification. However, the effects of lensing may still be 
detectable at much lower levels, in which case they might provide
some useful information upon parameters of the binaries. Moreover,
a failure to include lensing could possibly affect current methods 
of parameter estimation, significant given the importance of eclipsing
binaries in providing us with fundamental stellar data. 

I will show in this paper that the main impact of lensing is upon the
depth of secondary eclipse which can in turn influence the deduced
radii of the two stars. I begin in section~\ref{sec:deflect} by
outlining how one can include the effect of gravitational deflection
upon binary light curves; I obtain a simple estimate of the magnitude
of lensing amplification. In section~\ref{sec:simul} I present
simulations of eclipse light curves, and in section~\ref{sec:degen} I
discuss what effect lensing can have upon parameter determination. I
estimate the effects of lensing for known systems with white dwarfs
and discuss whether they will ever be of significance for black-holes
and neutron star binaries in section~\ref{sec:real}. 

\section{The effect of gravitational deflection of light}
\label{sec:deflect}
\begin{figure}
\hspace*{\fill}
\psfig{file=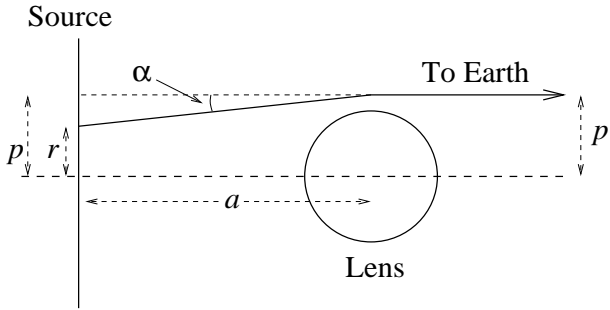,width=80mm}
\hspace*{\fill}
\caption{The figure shows the deflection of light from the companion
star (``source'') a distance $a$ from the compact object (``lens''). 
The light starts a distance $r$
from the point immediately behind the compact object as far as the observer
is concerned, is deflected by angle $\alpha$, and then appears to have 
originated $p$ from the sub-stellar point.}
\label{fig:deflect}
\end{figure}

The equations governing gravitational lensing by point sources have
been detailed in many papers (e.g. Blandford \& Narayan, 1992). 
However, in order to be self-contained and to develop the
equations in a form suitable for application to binary stars, in
this section I derive the relevant formulae. The geometry I will use
to describe this bending is illustrated in Fig.~\ref{fig:deflect}.
The phase illustrated corresponds to secondary eclipse, as the compact
object passes in front of its usually larger companion.  Most of the
deflection occurs close to the compact object and therefore I
approximate the path of the light by the two straight lines
shown. Distances are referred to the point on the companion star along
the line-of-sight through the centre of the compact object.  Light
starting a distance $r$ from this point is deflected by an angle
$\alpha$, and will appear to have originated a distance $p$ from
it. The distance $p$ is also the distance of closest approach to the
compact object and therefore
\[ \alpha = \frac{4 G M_1}{c^2 p}. \]
%
% addition
%
I assume here the standard deflection by a point mass which also
applies to the deflection by a spherically symmetric distribution
of mass contained within a radius less than the distance of closest
approach $p$ (and for example is the formula appropriate for deflection of
light by the Sun). This will be a very good approximation since the
white dwarfs I consider in this paper are either known or highly
likely to be rotating well below their break-up speeds.
Assuming that $\alpha \ll 1$, which will be the case if the radius
of the companion star $R_2 \ll a$, where $a$ is the orbital
separation,
\[ p = r + a \alpha.\]
This then leads to a quadratic in $p$ with solutions
\[ p = \frac{r \pm \sqrt{r^2+4R_e^2}}{2}, \]
where $R_e$, the Einstein radius, is given by
\[  R_e = \sqrt{\frac{4 G M_1 a}{c^2}}. \]
The latter expression is equivalent to the more commonly seen expression
\[  R_e = \sqrt{\frac{4 G M_1}{c^2}}
\sqrt{\frac{D_{sl} D_{ol} }{D_{sl}+D_{ol}}} \]
in the limit that the source-lens distance $D_{sl} = a$ is much less
than the observer-lens distance $D_{ol}$.
The negative root of $p$ corresponds to large deflections with the
light travelling below the centre of the lensing star in 
Fig.~\ref{fig:deflect}. It corresponds to the weaker of the two images 
in the standard microlensing analysis.

In the next section I present computations of light curves taking
account of the deflection of light and occultation by a white dwarf.
Before doing so, it is instructive to derive an estimate of the
change in flux from lensing that occurs when the stars are precisely
in line, as this serves to illustrate the likely importance of
lensing. For simplicity, I ignore any flux from the lensing star
itself. In this case, the companion star is imaged into two annuli, one for
each root of the quadratic equation. The positive root leads to 
a larger annulus than the negative root, but the two are exactly
joined at $p = R_e$. Assuming that the companion has uniform surface
brightness, and given that surface brightness is preserved during
lensing, then the two annuli appear as a single annulus with inner and
outer radii 
\begin{eqnarray*}
p_{\rm in}  &=& \frac{\sqrt{R_2^2 + 4R_e^2} - R_2}{2}\\
p_{\rm out} &=& \frac{R_2 + \sqrt{R_2^2 + 4R_e^2}}{2}.
\end{eqnarray*}
These radii result from putting $r = R_2$ for the negative and
positive roots respectively, and taking the modulus in the first case.
Given the preservation of surface brightness, the flux compared to 
the flux from the secondary alone is just the ratio of the visible areas:
\[ \frac{p_{\rm out}^2 - {\rm max}(R_1,p_{\rm in})^2}{R_2^2}. \]
The second term allows for occultation by the compact object.

Assuming, as will usually be the case for the binaries of interest in
this paper, that $R_e \ll R_2$ (the reverse of the situation for
microlensing between widely separated stars), then the above
expression can be approximated as
\[ \frac{R_2^2 + 2R_e^2 - {\rm max}(R_1,R_e^2/R_2)^2}{R_2^2}. \]
For white dwarfs, $R_1 = R_W \gg R_e^2/R_2$ and we have a flux ratio of
$(R_2^2 + 2R_e^2 - R_W^2)/R_2^2$ compared to 
$(R_2^2 - R_W^2)/R_2^2$ in the absence of lensing effects. 
For black-holes and neutron 
stars, whenever lensing is at all significant $R_1 < R_e^2/R_2$ 
(i.e. no occultation) and the amplification will be approximately
$(R_2^2 + 2 R_e^2)/R_2^2$.

In each case, the fractional increase compared to the case where
lensing is ignored is $2(R_e/R_2)^2$
From my assertion that $R_e \ll R_2$, this is evidently a small effect.
However, as I remarked earlier, the timing of the lensing, such that 
the increase coincides with secondary eclipse, effectively amplifies
its significance. The secondary eclipse (which only occurs at all in 
the white dwarf case) has fractional depth $(R_W/R_2)^2$, and
therefore, the lensing will reduce this by a
fraction
\[ f = 2 \left(\frac{R_e}{R_W}\right)^2 .\]
Thus for lensing to be significant in the context of the secondary
eclipse, we require that $R_e \sim R_W$ as opposed to $R_e \sim R_2$
as the earlier expression might have suggested. 

I will estimate values of $X = R_e/R_W$ and $f$ for known systems in 
section~\ref{sec:real}, but first I present simulated light curves 
demonstrating the lensing thus far described.

\section{Simulated light curves}
\label{sec:simul}
In order to demonstrate the influence of lensing upon eclipse light
curves, I have calculated some examples. I consider only the white
dwarf case since there is no eclipse in the case of black-holes and
neutron stars. I defined a polar grid over the projected (circular)
face of the source star, with uniform steps in radius and the number
of azimuths adjusted to keep the area per point approximately
constant. For each point, the value of $r$ is computed and then
compared with the value
\[r_{\rm min} = R_W (1 - X^2),\]
where $X = R_e/R_W$. This value of $r$ corresponds to $p = R_W$; 
for $r < r_{\rm min}$, the point is occulted by the white dwarf. 

As before, the lensing amplification can be worked out from the amplification
of areas since surface brightness is preserved. This is given by
\begin{eqnarray*}
A &=& \frac{p \,dp}{r\,dr} \\
  &=& \frac{1}{4}\left(1 + \sqrt{1+4R_e^2/r^2}\right)
      \left(1+1/\sqrt{1+4R_e^2/r^2}\right) 
\end{eqnarray*}
This only accounts for the contribution from the positive root 
of the quadratic of the previous section. In general the negative root
needs to be included too, but not if $R_e < R_W$ since it is
then occulted by the white dwarf. I will assume that this is the
case, which also prevents numerical difficulties associated with the
infinite amplification of $r = 0$ (although as shown in the previous section, 
the integrated amplification remains finite). I will show in 
section~\ref{sec:real} that $X  = R_e/R_W < 1$ in all systems known where 
the lensing effect is potentially observable, although systems
violating this condition may one day be found. 

The light curves are then obtained by summing over all the visible
grid points, including the magnification appropriate for each one.  It
is straightforward within this scheme to include limb darkening.  Limb
darkening is useful in standard microlensing work because it modifies
the light curves in ways that can lead to new information
(Valls-Gabaud, 1998;  Han, Park \& Jeong, 2000). It is also possible
to use caustic crossing events to measure limb darkening
(Albrow et al., 1999). In these cases limb darkening breaks the
degeneracy between models that would otherwise predict similar
lightcurves. However limb darkening seems much less likely to help in
the case of self-lensing considered here, because its influence upon
the light curve is to a large extent degenerate with alterations in
the radii of the stars. In section~\ref{sec:degen} I will show that
lensing suffers the same degeneracy and therefore limb darkening is
likely to make it harder rather than easier to detect the effects of
lensing. For this reason, and in order to emphasize the effects of
lensing which have not been included in eclipsing binary light-curve
analyses before, I do not include limb darkening in any of the simulations 
presented here.

\begin{figure}
\hspace*{\fill}
\psfig{file=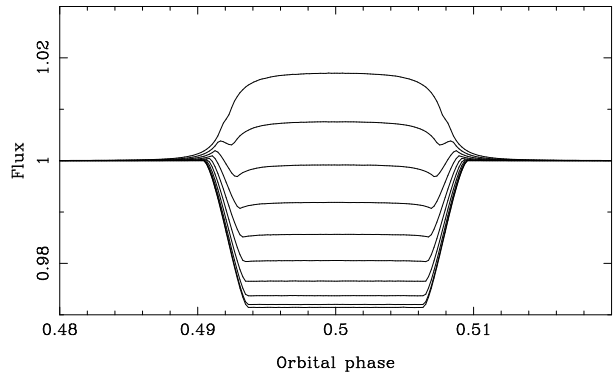,width=80mm}
\hspace*{\fill}
\caption{Simulated light curves at the (total) secondary eclipse of a binary 
with $R_1/a = 0.01$, $R_2/a = 0.05$, $i = 90^\circ$ for $X = R_e/R_W = 0.0$ 
(bottom curve) to $0.9$ (top curve) in steps of $0.1$.}
\label{fig:sim1}
\end{figure}

\begin{figure}
\hspace*{\fill}
\psfig{file=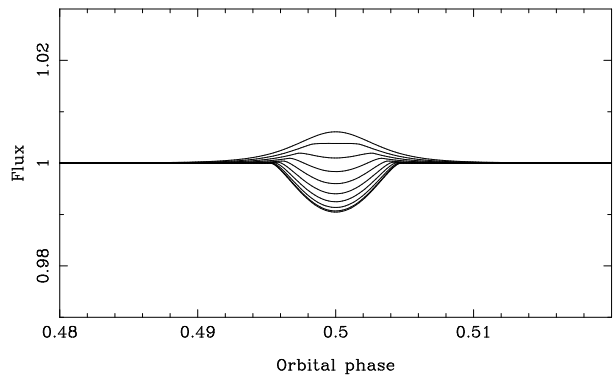,width=80mm}
\hspace*{\fill}
\caption{Simulated light curves at the (partial) secondary eclipse of a 
binary with $R_1/a = 0.01$, $R_2/a = 0.05$, $i = 87^\circ$ for 
$X = R_e/R_W = 0.0$ (bottom curve) to $0.9$ (top curve) in steps of 
$0.1$.}
\label{fig:sim2}
\end{figure}

First I consider a binary
with $R_1/a = 0.01$ and $R_2/a = 0.05$ (e.g. a white dwarf and a very
low mass M dwarf/brown dwarf). Light curves around the secondary 
eclipse for such a binary for $X = 0.0$ to $0.9$ in steps of $0.1$ are 
plotted in Figs.~\ref{fig:sim1} and \ref{fig:sim2} which show total
and
partial eclipses ($i = 90^\circ$ and $87^\circ$ respectively).

As predicted the mid-eclipse depth is reduced to almost zero for 
$X = 0.7 \approx 1/\sqrt{2}$. Gravitational lensing is also visible 
as an increase above the out-of-eclipse flux just before and after 
eclipse for lower values of $X$, and at all stages 
of eclipse for $X > 0.7$. Given the small depth of eclipse in 
Figs.~\ref{fig:sim1} and \ref{fig:sim2}, and limitations of signal-to-noise,
the overall effect is likely to be undetectable for $f < 0.1$, or $X < 0.2$.

\begin{figure}
\hspace*{\fill}
\psfig{file=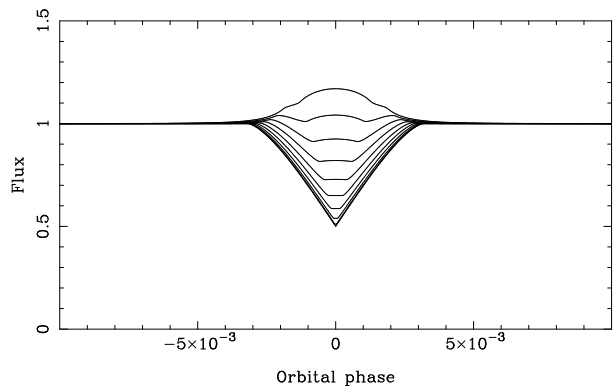,width=80mm}
\hspace*{\fill}
\caption{Simulated light curves at the primary eclipse of a 
binary with $R_1/a = R_2/a = 0.01$ (i.e. two white dwarfs) and 
$i = 90^\circ$ for $X = R_e/R_W = 0.0$ (bottom curve) to $0.9$ 
(top curve) in steps of $0.1$.}
\label{fig:sim3}
\end{figure}

Fig.~\ref{fig:sim3} shows the primary eclipse of a double-degenerate
eclipsing binary consisting of two identical white dwarfs with $R_1/a
= R_2/a = 0.01$. Although the two stars have the same radius, the
gravitational deflection causes the eclipse to appear total
(flat-bottomed section) because it makes the more distant white dwarf
appear larger. This will also be the case during the other eclipse,
and so we can have the curious situation where {\em both} eclipses are
transits of one star in front of an apparently larger star. The
overall effect may appear similar in magnitude to the earlier
simulations, however the eclipse depth is now much larger (note the
very different vertical scales), and therefore the changes are more
significant compared to the out-of-eclipse flux than before.

\begin{figure}
\hspace*{\fill}
\psfig{file=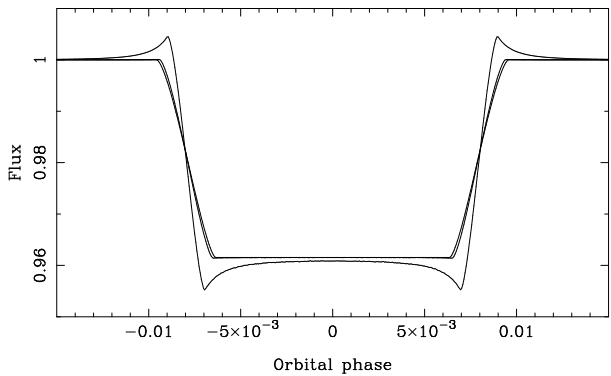,width=80mm}
\hspace*{\fill}
\caption{The figure shows the secondary eclipse for a binary
with $R_1/a = 0.01$, $R_2/a = 0.05$, $i = 90^\circ$ and $X = 0.3$
and $0.6$, compared to the shape of primary eclipse; the secondary eclipses
have been shifted and scaled to aid the comparison.}
\label{fig:sim4}
\end{figure}

At low values of $X$, the main effect of the gravitational bending is
to reduce the depth of the secondary eclipse. I will show in the next
section that this cannot be distinguished from a change in the radius
ratio of the two stars. Therefore this is an effect that should be
included as part of light curve modelling rather than as a free parameter
of a fit. The lensing does have other effects that could not be
mimicked by alterations in other parameters of the binary. These are
illustrated in in Fig.~\ref{fig:sim4} in which the secondary and
primary eclipse shapes are compared.
Lensing steepens the slope of the ingress and egress of the secondary
eclipse when compared to (and scaled to the same depth as) primary
eclipse. The steepening is easily understood: the start of ingress
(first contact) is delayed as we carry on seeing the companion to the
white dwarf owing to the light deflection. In a similar manner, once
past the half-way point between first and second contact, the
deflection acts to advance the time at which we first see the white
dwarf within the boundary of its companion (second contact). 
Unfortunately, in absolute terms this effect is very small
until $X > 0.7$ (Fig.~\ref{fig:sim1}), by which time lensing should
be obvious from a complete absence of eclipse. 

Even if lensing does affect the secondary eclipse significantly,
does it matter in the overall scheme of things, given that, as I have
shown, it is small in absolute terms? To address this, in the
next section I present a brief digression upon how light curves
of eclipsing binary stars are used to measure the radii of the 
stars and how one can avoid degeneracy in the solution.

\section{Degeneracy in determination of eclipsing binary parameters}
\label{sec:degen}
Eclipsing binary stars are the source of the most precise measurements
of masses and radii of stars other than the Sun. Photometry alone can
lead to the determination of the radii of the two stars (scaled by the
orbital separation) and the orbital inclination; spectroscopy can then
lead to a complete solution for masses and radii. However, for
white dwarf eclipsing binaries, it is usually much easier to see the 
primary eclipse and this is what observers tend to concentrate upon.
In this case, there is no unique solution without invoking less direct
assumptions.

The radii of stars in eclipsing binaries are constrained by the lengths 
and shapes of the eclipses. Most of the available information is conveyed
by the contact phases. For two spherical stars of radii $R_1$ and $R_2$
in a binary of separation $a$ and orbital inclination $i$, it can be
shown that:
\begin{eqnarray*}
\frac{R_1}{a} &=& \frac{1}{2}\left[(\cos^2 i + \sin^2 i \sin^2 \phi_1)^{1/2}\right. \\
              & & \;\;\left. - (\cos^2 i + \sin^2 i \sin^2 \phi_2)^{1/2}\right] \\
\frac{R_2}{a} &=& \frac{1}{2}\left[(\cos^2 i + \sin^2 i \sin^2 \phi_1)^{1/2}
\right.\\
& &\;\;\left.+ (\cos^2 i + \sin^2 i \sin^2 \phi_2)^{1/2}\right]
\end{eqnarray*}
where $\phi_1$ and $\phi_2$ are the first and second contact phases (expressed
in radians). The contact phases are measured from the light curves; $R_1/a$,
$R_2/a$ and $i$ are unknowns. With only two relations (the third and fourth 
contacts give no further information), the problem is degenerate. The
depth of eclipse does not help as this just gives the ratio of surface
brightnesses of the two stars (for any particular inclination).
The problem is illustrated
\begin{figure}
\hspace*{\fill}
\psfig{file=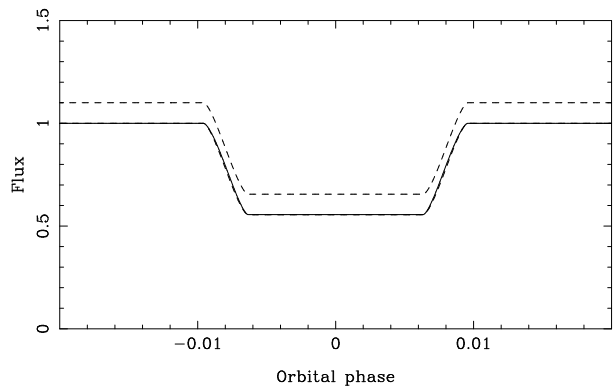,width=80mm}
\hspace*{\fill}
\caption{The solid line shows the primary eclipse of a binary with
$R_1/a = 0.01$, $R_2/a = 0.05$ and $i = 90^\circ$. The dashed line,
plotted twice, in one case displaced upwards for clarity, shows
the primary eclipse of a binary with $R_1/a = 0.00816$, 
$R_2/a = 0.0612$ and $i = 88^\circ$. No limb darkening was included.
}
\label{fig:degen}
\end{figure}
in Fig.~\ref{fig:degen} which shows the eclipse of two binaries with
very different relative radii and inclinations, but adjusted to give the
same contact phases and eclipse depth. The two eclipses are almost
indistinguishable on the plot, and would be utterly indistinguishable
in practice because of noise.

The degeneracy can be lifted by detection of the secondary eclipse
since this gives us $(R_2/R_1)^2$, given that we know the surface
brightness ratio from the primary eclipse (and ignoring limb darkening
which introduces an inclination dependence as well). For instance this
parameter has values of $0.040$ and $0.018$ for the two light curves
of Fig.~\ref{fig:degen}, and even relatively crude measurements would
distinguish the two models.  It is not often possible to detect the
secondary eclipse in white dwarf binaries (I discuss two cases where
it is in the next section), but given its usefulness and the growing
number of these systems (Marsh, 2000), more suitable systems
will be found in the future.

In summary, although lensing effects are weak, they affect exactly the
part of the light curve -- the secondary eclipse -- that is needed to obtain reliable
measurements of the radii.  We now estimate the magnitude of the effect
for some known systems.

\begin{table*}
\caption{Values of the lensing parameter for some known binary stars.}
\begin{tabular}{lllllllll}
Name      & $M_W$     & $M_2$   & Type & $P$         & $X$     & $f$  & 
Eclipses? & Reference(s)\\
          & $\mathrm{M}_\odot$& $\mathrm{M}_\odot$& &d&$R_e/R_W$&$2X^2$&& \\
\hline
RR~Cae        & $0.365$  & $0.089$ & M dwarf   & $0.304$  & $0.12$& $0.03$& 
Yes & Bruch \& Diaz (1998)\\
V471~Tau      & $0.72$   & $0.70$  & K dwarf   & $0.521$  & $0.38$& $0.29$& 
Yes & Young \& Lanning (1975)\\
GK~Vir        & $0.51$   & $0.10$  & M dwarf   & $0.344$  & $0.19$& $0.07$&
Yes & Fulbright et al. (1993) \\
NN~Ser        & $0.57$   & $0.12$  & M dwarf   & $0.130$  & $0.16$& $0.05$&
Yes & Catalan et al. (1994)\\
KPD 0422+5421 & $0.53$   & $0.51$  & sdB star  & $0.090$  & $0.14$& $0.04$& 
Yes & Orosz \& Wade (1999)\\
KPD 1930+2752 & $0.97$   & $0.50$  & sdB star  & $0.095$  & $0.34$& $0.23$& 
Probably & Maxted, Marsh \& North (2000)\\
PG 0940+068   & $0.63$   & $0.50$  & sdB star  & $8.33$   & $0.80$& $1.28$& 
Unknown & Maxted et al. (2000)\\
L870-2        & $0.52$   & $0.47$  & white dwarf& $1.556$ & $0.37$& $0.27$&
\parbox[t]{0.5in}{Probably not.} & 
\parbox[t]{1.2in}{\raggedright Saffer, Liebert \& Olszewski (1988), Bergeron et al. (1989)}
\end{tabular}
\label{tab:real}
\end{table*}

\section{The significance of lensing in known systems}
\label{sec:real} 
\subsection{White dwarf binary stars}
The magnitude of the effect that the gravitational deflection of light 
has upon light curves of eclipsing white dwarf binaries depends upon
the parameter
$X$ introduced in the section~\ref{sec:deflect}. Using Kepler's
third law, this can be expressed as
\begin{eqnarray*}
 X & = & 0.59 \left(\frac{M_W}{{\rm M}_\odot}\right)^{1/2}
\left(\frac{M_W+M_2}{{\rm M}_\odot}\right)^{1/6} \times \\
   &   &\;\;\;\;\;\;\;\left(\frac{R_W}{0.01\,{\rm R}_\odot}\right)^{-1}
        \left(\frac{P}{1\,{\rm d}}\right)^{1/3},
\end{eqnarray*} 
where $M_2$ is the mass of the white dwarf's companion and $P$ is the
orbital period of the binary (a circular orbit is assumed). Assuming 
Nauenberg's (1972) analytic mass-radius relation for white dwarfs
with an electron molecular mass $\mu_e = 2$, appropriate for He or
CO white dwarfs:
\[ \frac{R_W}{{\rm R}_\odot} = 0.01125 
\left(\left(\frac{M}{1.454\,{\rm M}_\odot}\right)^{-2/3}-
      \left(\frac{M}{1.454\,{\rm M}_\odot}\right)^{2/3}\right)^{1/2}. \]
I list values of $X=R_e/R_W$ and $f = 2 X^2$ for some known systems in 
Table~\ref{tab:real}. The secondary eclipses are visible in KPD~0422+5421 
(Orosz \& Wade, 1999) and
in RR~Cae (Bruch \& Diaz, 1998), although the latter requires 
confirmation. The sixth system, KPD~1930+2752, is very probably eclipsing
(Maxted, Marsh \& North, 2000), 
although in this case only the transit is visible, and like RR~Cae, requires 
confirmation. The last two systems listed are not known to be eclipsing, 
although PG~0940+068, an sdB star which probably has a white dwarf companion,
shows large radial velocity variations indicative of high inclination
(Maxted et al., 2000). 

The values of $f$ are large in several cases, but these are systems
which either do not eclipse or in which the secondary star is so much
larger than the white dwarf that it will be hard to detect the transit
from the ground. On the other hand it is clear that $f$ can be very
significant in systems such as PG~0940+068 where there is nothing to
prevent detection of eclipses, provided that they occur.

There is one system, KPD~1930+2752, for which all the conditions are
appear correct for significant lensing effects. If the transit of
the white dwarf in this system is confirmed 
(Bill\`eres et al., 2000; Maxted, Marsh \& North, 2000), it is $\sim 25$\% less
deep that it would be in the absence of lensing. Thus neglect of lensing
in this system would lead to $\sim 12$\% underestimate of the radius of
the white dwarf and a consequent over-estimate of its mass. This is significant
because the total mass of this system determines its status as a possible
Type~Ia progenitor and yet at the moment it relies on an assumed mass of
$0.50\,\mathrm{M}_\odot$ for the sdB star. Measurement of the eclipse depth is probably the
best way to avoid this assumption.

\subsection{Neutron star and black-hole binaries}
Neutron star and black-hole binaries will not show any secondary
eclipse, but will still display the $2(R_e/R_2)^2$ fractional
increase derived in section~\ref{sec:deflect}. A measurement of this 
increase could be used to measure the mass of the compact star, a
parameter of considerable interest for both types of compact object.
In what systems can we hope to see this effect? 

Gould (1995) concluded that millisecond pulsar binaries (pairs of
neutron stars) show the most promise. To this I add millisecond
pulsars with white dwarf companions (e.g. van Kerkwijk, Bergeron \& Kulkarni, 1996). For a
$1.4\,{\rm M}_\odot$ neutron star lensing a $0.6\,{\rm M}_\odot$ white
dwarf in a binary of orbital period $1$ day, $R_e = 0.004\,{\rm
R}_\odot$, and a $\sim 30$\% increase in flux is possible, for an
edge-on system. Suitable systems could be identified from any
detectable Shapiro delay in the pulses from the neutron star. The
problem is that, as Gould discussed, the chance of a binary having the
right orientation is $\approx R_e/a$, which is of order 1 in several
hundred, and it is not clear that we will ever find enough of these
systems to find one with significant lensing. This may still be true
even if the conditions were relaxed to allow for smaller
amplifications. The problem here, in contrast to the white dwarf
binaries, is that most of these systems are faint ($V > 20$), and
so $0.01$ magnitude or smaller effects are going to be difficult
to detect.

The X-ray transients also contain neutron stars and black-holes, and
in their low states are dominated by the light of the companion to the
compact object which fills its Roche lobe. The increase in light from lensing
can be written
\[ \frac{8 G M_1}{c^2 a (R_2/a)^2} \]
where the mass ratio $q$ is defined as $q = M_2/M_2$ and 
where $R_2/a$ can be obtained from the mass ratio $q = M_2/M_1$ and 
Roche geometry:
\[ \frac{R_2}{a} = \frac{0.49 q^{2/3}}{0.6q^{2/3} + \ln(1+q^{1/3})},
\]
Eggleton (1983).

Some example parameters of black-hole candidates are: GRO J1655-40,
$M_1 = 7\,{\rm M}_\odot$, $q = 0.333$, $P = 2.6\,{\rm d}$ (Orosz \& Bailyn, 1997); 
A0620-00, $M_1 = 10\,{\rm M}_\odot$, $q = 0.067$, $P =
0.323\,{\rm d}$ (McClintock \& Remillard, 1986; Marsh, Robinson \& Wood, 1994; Shahbaz, Naylor \& Charles, 1994); 
QZ~Vul, $M_1 = 8.5\,{\rm M}_\odot$, $q = 0.04$,
$P = 0.344\,{\rm d}$ (Harlaftis, Horne \& Filippenko, 1996).  None of these are seen
edge-on, but even if they were, the lensing increase would amount to
only one milli-magnitude or less in each case.  It is hard to believe
that such small increases will ever be detectable against the
intrinsic variability that these systems show even in their low states; the
same applies to accreting neutron stars which have the disadvantage of
lower masses as well. The option of finding systems with much larger
separations is prevented by the need for mass transfer to occur in
order that these systems are found in the first place.

\section{Discussion}
Evidently gravitational lensing makes little impact upon the light
curves of most eclipsing binaries. This can be put down to the
requirement for the large separation needed to make the Einstein
radius $R_e = \sqrt{4 G M a}$ as large as possible, but which also
makes the chances of eclipse small. For example, if a
double-degenerate binary has $M_1 = M_2 = 0.6\,{\rm M}_\odot$, so that
$R_1 = R_2 = 0.0125 \,{\rm M}_\odot$, then for $X > 0.4$, say, we need
$a > 5 \,{\rm R}_\odot$.  The chance of the binary eclipsing is
$(R_1+R_2)/a$, which amounts to about 1 in 200. For a white dwarf/M
dwarf binary, with $R_2 = 10 R_1$, this becomes 1 in 40. Although
small, this is a considerably higher fraction than suggested by
Gould's (1995) statement that $>10^5$ observations of each of
$10^4$ white-dwarf binaries would be needed to find one example. The
difference results from my focus upon events strong enough to affect
light curve analysis, but still much feebler than those considered by
Gould: if one is interested in 3 milli-magnitude effects as opposed to 0.3
magnitudes, since $f = 2X^2 \propto a$, the probability of eclipse is 100
times greater. This brings Gould's $10^4$ binaries down to 100; in
addition, the $10^5$ observations per system are unlikely to be
needed because most methods of detecting white dwarf binaries also
give the orbital phase, and thus one knows when to look for eclipses
and lensing.

This discussion is given weight by the identification of one system,
KPD~1930+2752, for which lensing has a large enough effect that it
should be included in light curve models (subject to verification of the
white dwarf transit in this system). It is very likely that other such
systems will be found in the near future.

\section{Conclusions}
I have estimated the effects of gravitational lensing within eclipsing
binaries containing white dwarfs. Although these are usually small, 
since lensing is maximum at conjunction, it has the
potential to reduce the depth of secondary eclipse as the white dwarf
transits its companion; measurement of this depth is essential in
deriving precise radii of the stars. I find that the eclipse depth is
reduced by a fraction $f = 2 X^2$, where $X = R_e/R_W$ and the
Einstein radius $R_e = \sqrt{4G M_W/c^2}$. Estimates of $f$ for
some known binaries are less than a few percent, but there is one sdB/white
dwarf binary, KPD~1930+2752, which shows signs of a transit of the
white dwarf which lensing should affect by $\sim 25$\%. More such
systems are sure to be found in the future.

I also consider binaries with black-hole or neutron star components.
Lensing provides a potential method of measuring the mass of the
compact object in such binaries. However, I agree with
Gould's (1995) conclusion that their rarity (and in the case
of semi-detached binaries, their intrinsic variability) make it
unlikely that lensing will prove a useful tool in practice.

\section*{Acknowledgments}
Use was made of the computational facilities of the Southampton node 
of the UK STARLINK computer network.

\end{document}